\documentclass[12pt]{iopart}
\usepackage{cite}
\usepackage[pdftex]{graphicx}
\usepackage{moreverb}
\usepackage{amssymb}                    
\usepackage{amscd}                   
\usepackage{wasysym} 
\usepackage{makeidx} 
\newcommand{\ML}{\textsc{Matlab}\ }
\newcommand{\re}{{\rm e}}
\newcommand{\ri}{{\rm i}}
\begin{document}
\jl{1}
\title[Computations in quantum mechanics made easy]
{Computations in quantum mechanics made easy}
\author{H J Korsch$^{1}$ and K Rapedius$^{2}$}

\address{$^{1}$ FB Physik, Universit\"at Kaiserslautern, D-67653
Kaiserslautern, Germany}
\address{$^{2}$ Karlsruhe Institute of Technology (KIT), Adenauerring 2, D-76131 Karlsruhe, Germany}

\ead{korsch@physik.uni-kl.de, kevin.rapedius@kit.edu},

\begin{abstract}
Convenient and simple numerical techniques for performing quantum computations based on 
matrix representations of Hilbert space operators are presented and illustrated by various examples.  
The applications include the calculations of spectral and dynamical properties for one-dimensional 
and two-dimensional single-particle systems as well as bosonic many-particle and open quantum 
systems. Due to their technical simplicity these methods are well suited as a tool for teaching 
quantum mechanics to undergraduates and graduates. Explicit implementations of the presented 
numerical methods in \ML are given.
\end{abstract}

\submitto{\EJP}

%

\section{Introduction}
\label{s-int}
In \cite{02computing} it was shown how to calculate the spectra of one-dimensional quantum 
systems in a simple, convenient and effective way by means of matrix representations of Hilbert 
space operators. Here we extend these techniques in various ways including the computation of 
dynamical properties as well applications to higher dimensional systems,  bosonic  many-particle 
and open quantum systems.

As discussed in \cite{02computing} the basic building blocks for the discrete matrix representation 
of operators used in the following programs are the operators $\hat a$ and $\hat a^\dagger$, well 
known from the harmonic oscillator, where they act as ladder operators on the
harmonic oscillator eigenstates $|n\rangle$, $n=0,\,1,\,2,\ldots$\,:
\begin{eqnarray}
\label{aad}
\hat a \,|n+1\rangle =\sqrt{n+1}\,|n\rangle\,,\ 
\hat a^\dagger |n\rangle =\sqrt{n+1}\,|n+1\rangle\,,\ 
\hat a^\dagger \hat a \,|n\rangle =n\,|n\rangle\,.
\end{eqnarray}
Motivated, e.g., by the application to the radiation field described by 
harmonic oscillators with frequency $\omega_0$ these operators create
or annihilate a photon of this frequency or, more generally, a  bosonic
particle in second quantization. Therefore these operators are  also known as
creation and annihilation operators. Here we will mainly use the matrix
representation of these operators in the harmonic oscillator basis:
\begin{eqnarray}
\label{amatrix}
{\hat a}=
\left(\begin{array}{ccccc} 
0  &  \sqrt{1}  &  0  &  0  & \ldots\\
0 & 0 & \sqrt{2} & 0 & \ldots\\
0 & 0 & 0 & \sqrt{3} & \ldots\\
0  &  0  &  0  &  0  & \ldots\\
\vdots &\vdots&\vdots&\vdots& \ddots\\
\end{array}\right).
\end{eqnarray}
As an introduction to the numerical applications presented below, the 
following  short \ML code shows the construction of these matrices and
their application to a basis vector
{\small
\begin{listing}[1000]{1001}       
N = 4;  Np=N+1; n = 1:N; 
a = diag(sqrt(1:N),1); ad=a';
n = [0 0 1 0 0]';
a*n; ad*n; ad*a*n
\end{listing}
}
\noindent
where the properties (\ref{aad}) can be tested.
It should be noted, however, that such numerical matrix representations are 
necessarily finite, which  causes numerical  errors. The reader may try for example
to test the validity of the commutation relation $[\hat a,\hat a^\dagger]=1$
numerically or to construct the eigenstates of the annihilation operator $\hat a$,
the coherent states, numerically. 
In many cases these  finite size errors can be reduced  by increasing the matrix size. 

In addition, we will use the representation of the position and momentum
operators 
\begin{eqnarray}
\label{xpmatrix}
\hat x=\case{s}{\sqrt{2}}\,\big(\hat a^\dagger+\hat a\big)\ ,\quad
\hat p=\case{\rmi}{s\sqrt{2}}\,\big(\hat a^\dagger-\hat a\big)
\end{eqnarray}
where the scaling parameter is chosen as $s=1$ in the
following (see \cite{02computing} for details).
These matrix representations of operators can now be used to construct
in a simple way matrix representations of other operators, 
such as the Hamiltonian or time-evolution operators.

This paper is organized as follows: In section \ref{s-one} we briefly review the calculation of
quantum eigenvalues for one-dimensional systems \cite{02computing} and extend the analysis to 
time-dependent calculations. The example applications include Bloch-oscillations in a tilted periodic
lattice. Section \ref{s-ang} illustrates the use of matrix representation for angular momentum 
operators. In section \ref{s-two} it is shown how to calculate the spectrum of a two-dimensional 
quantum system. Applications to  bosonic many-particle systems are illustrated in section \ref{s-many} by 
means of the Bose-Hubbard model. Finally, the dynamics of an open quantum system described 
by a Lindblad master equation is calculated in section \ref{s-lindblad}. 

%
\section{One-dimensional systems}
\label{s-one}
\begin{figure}[htb]
\begin{center}
\includegraphics[width=80mm,clip]{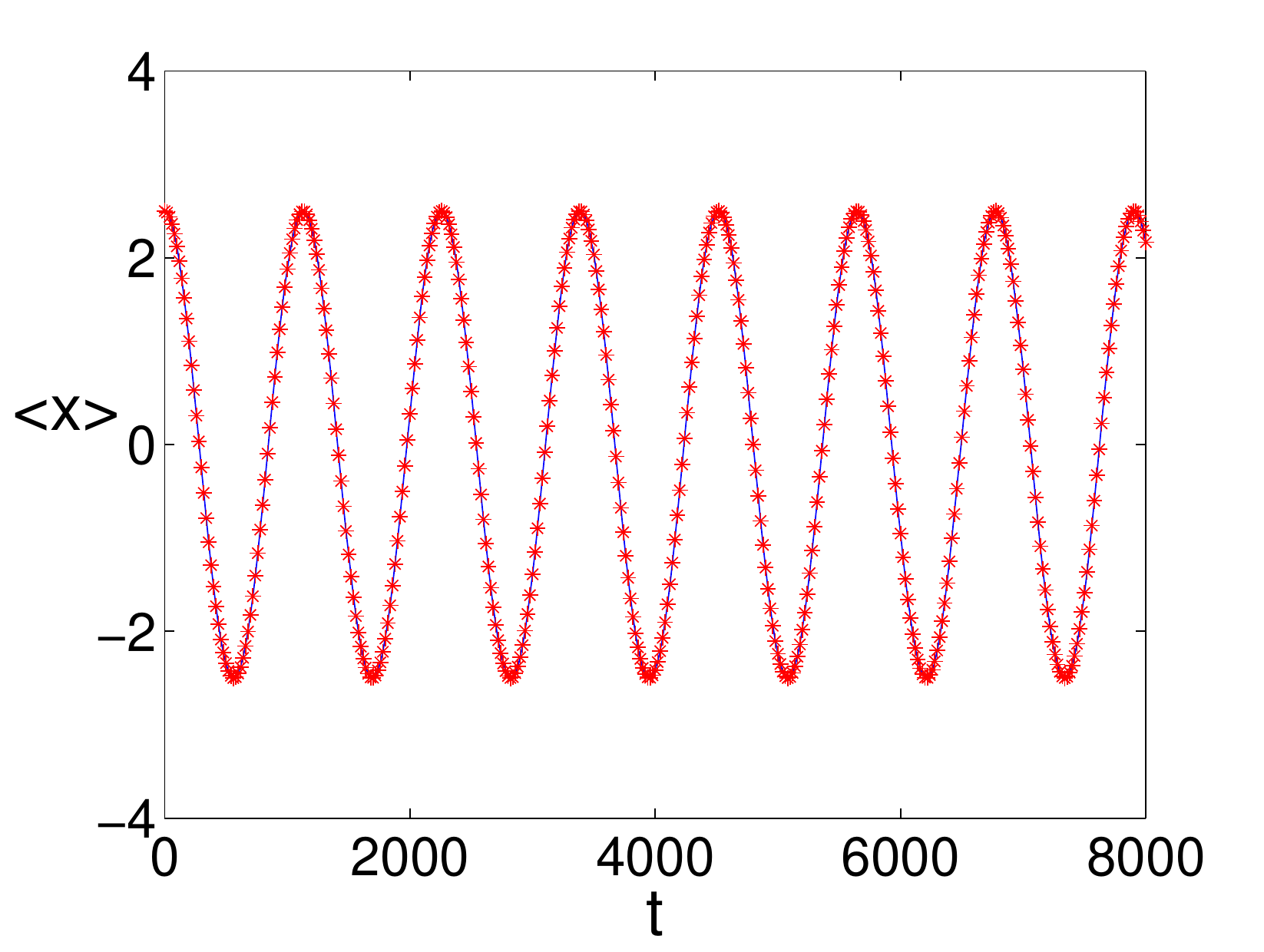}
\caption[ ]{Double well potential: Expectation value $\langle \hat x \rangle$ as a function of time for a Gaussian wave packet initially at the right well (blue line) compared to the 
approximation  $x(t)$ (red stars).\label{f-eigendw2} }
\end{center}
\end{figure}

As a first example we consider the calculation of bound state energies for a
single particle in  a one-dimensional potential described by the  
Hamiltonian 
\begin{eqnarray}
\label{ham1}
\hat H=\frac{\hat p^2}{2m}+V(\hat x)\,,
\end{eqnarray}
which has already been discussed in \cite{02computing} for harmonic and 
quartic potentials. There a simple program code can be found for computing
the eigenvalues and, in addition, also the corresponding wave functions. As an
introduction we also list this code here, however for a simple double-well
potential
\begin{eqnarray}
\label{dwpot}
V(x)=\frac12\big(|x|-x_0)^2\,,
\end{eqnarray}
a potential with harmonic minima at $\pm x_0$ separated by a barrier
of height $x_0^2/2$ at $x=0$. 
In the program code
{\small
\begin{listing}[1000]{1001}
N = 100;  n = 1:N-1; x0 = 2.5;
m = sqrt(n);
x =  1/sqrt(2)*(diag(m,-1)+diag(m,1));
p = i/sqrt(2)*(diag(m,-1)-diag(m,1));
H = p^2/2+(sqrtm(x^2)-x0*eye(N))^2/2;
E = sort(eig(H)); E(1:8) 
\end{listing}
}
\noindent
where $\hbar$ and the particle mass $m$ are chosen as unity and
$x_0=2.5$. In addition it should be noted that the matrix function $|x|$ has been 
generated by taking the square \verb?x^2? followed by  the \ML matrix function 
\verb?sqrtm(x^2)?. The output of the program is a listing
of the first eight energy eigenvalues $E_n$, $n=0,\,\ldots,7$, where the
eigenvalues below the potential barrier are arranged in doublets. Let us have
a brief look at the time dynamics for an initial state chosen as the
ground state of the right potential well. This can be achieved by adding the lines 
{\small
\begin{listingcont}
H1 = p^2/2 + (x-x0*eye(N))^2/2;
[V,Eig] = eig(H1);
psi = V(:,1);
tstep = 20; t(1) = 0; xav(1) = psi'*x*psi;
U = expm(-i*H*tstep);
for nt=1:400
  t(nt+1) = t(nt)+tstep;
  psi = U*psi;
  xav(nt+1) = psi'*x*psi;
end
plot(t,xav); set(gca,'Fontsize',20)
xlabel('t','FontSize',24); ylabel('<x>','Rotation',0,'FontSize',24);
hold on
plot(t,x0*cos((E(2)-E(1))*t),'*r')
\end{listingcont}
}
\noindent
to the program above. Then the time evolution operator
$\hat U=\exp{-\rmi \hat H \Delta t/\hbar}$ is used to propagate the  
wave function over a sequence of $80$ time steps $\Delta t=20$.
At each step the expectation value $\langle \hat x\rangle$ is
computed and finally plotted as a function of time in comparison with
the approximation $x(t)=x_0\cos((E_1-E_0)t)$ as shown in figure \ref{f-eigendw2}. 

As a second application let us consider an extended one-dimensional system,
a particle in a periodic potential $V_0(x+d)=V_0(x)$ accelerated by a constant force $F$:
\begin{eqnarray}
\label{Vbloch}
V(x)=V_0(x)+Fx
\end{eqnarray}
In the tight-binding approximation, the Hamiltonian is expressed in terms
of the Wannier states $|n\rangle$ of the lowest band of the periodic 
potential, which are localized at the potential minima numbered by $n$.  
Taking only transitions 
between neighboring wells into account, the tight-binding Hamiltonian
reads
\begin{eqnarray}
\label{tbham}
\hat H= \sum_n (\epsilon+ dFn)|n\rangle \langle n| -\frac{\Delta}{4}\big(\,|n+1\rangle\langle n|
+|n\rangle\langle n+1|\,\big)
\end{eqnarray}
(see, e.g., \cite{04bloch1d} and references therein), where $\epsilon$ is the band energy and
$\Delta$ is the bandwidth.  For $F=0$ we have an almost free motion with quasimomentum 
$\kappa$  for energies
inside the band $E(\kappa)=\epsilon +\case{\delta}{2}\cos \kappa $, whereas for $F\ne 0$ the particle is,
contrary to a naive expectation, confined to a finite region in space performing
an oscillatory motion with a period $T_B=2\pi\hbar/dF$, the so-called
Bloch period. This is closely related to the fact that the
Hamiltonian possesses the equidistant eigenvalues 
$E_n=\epsilon+dFn$,  $n=0,\pm1,\pm2,\ldots$\,.  This Bloch
oscillation is illustrated by the following program for 
$F=0.005$, $d=2\pi$, $\Delta=1$, $\hbar=1$ and an array extending from $n=-60$
to $n=60$.  
{\small
\begin{listing}[1000]{1001}
d = 2*pi; F = 0.005; Delta = 1;
nmax = 60; nmin = -60;
n = nmin:nmax; nn = length(n);
m = ones(1,nn-1); 
H = d*F*diag(n,0)+(Delta/4)*(diag(m,-1)+diag(m,1));
\end{listing}
}
\noindent
The matrix representation of the Hamiltonian (\ref{tbham}) in the basis $|n\rangle$
is tridiagonal. Its construction is very similar to the
one used for the $x$ matrix in the program above. Here, however, there are
also non-vanishing entries along the diagonal and the other matrix elements
are constant. The interested reader
can easily calculate the eigenvalues of this Hamiltonian numerically and compare with
the analytical result given above. This will show that the results agree with the
exception of the regions close to the boundary of the finite $n$-array, where
the numerical values are not yet converged. In the following we will 
study the time evolution of an initial wave packet. 
We first consider a wave packet, localized initially at site $n=0$, which is
propagated over $N=2$ Bloch periods, each one is discretized by $J=80$
time steps: 
{\small
\begin{listingcont}
% initial state (gaussian)
psi = 0*n; psi(-nmin+1) = 1; % initial state (localized)
% time propagation (N Bloch periods, J steps each)
J = 80; N = 2; 
Psi = zeros(nn,N*J+1);
Psi(:,1) = psi; 
U = expm(-i*H*2*pi/d/F/J);
for nt = 1:N*J
    Psi(:,nt+1) = U*Psi(:,nt);
end
t = 0:N*J; t = t/J;
imagesc(t,n,abs(Psi))
set(gca,'ydir','normal','FontSize',20)
xlabel('t/T_B','FontSize',20)
ylabel('n','rotation',0,'FontSize',20)
\end{listingcont}
}
\begin{figure}[htb]
\begin{center}
\includegraphics[width=75mm,clip]{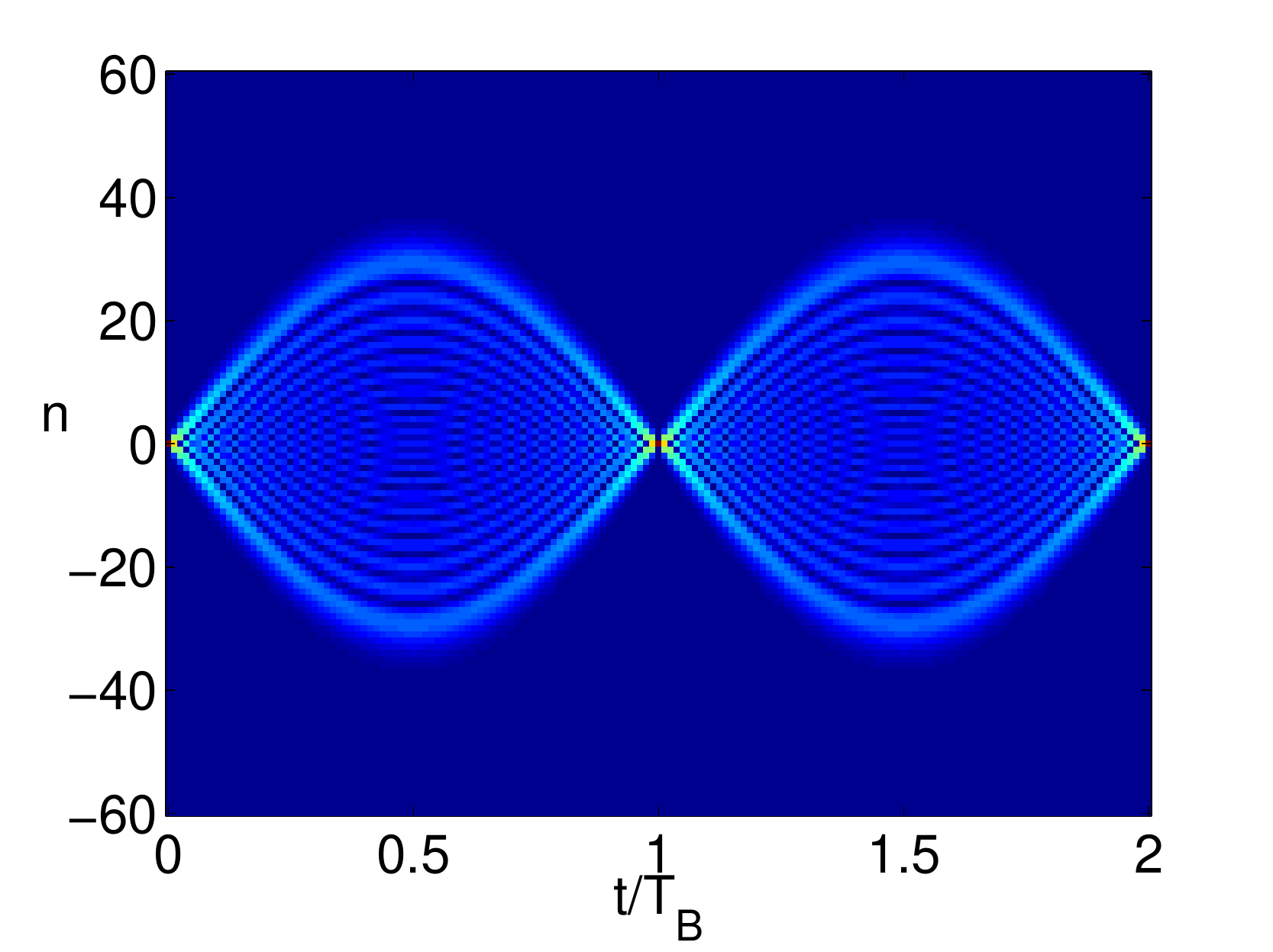}
\includegraphics[width=75mm,clip]{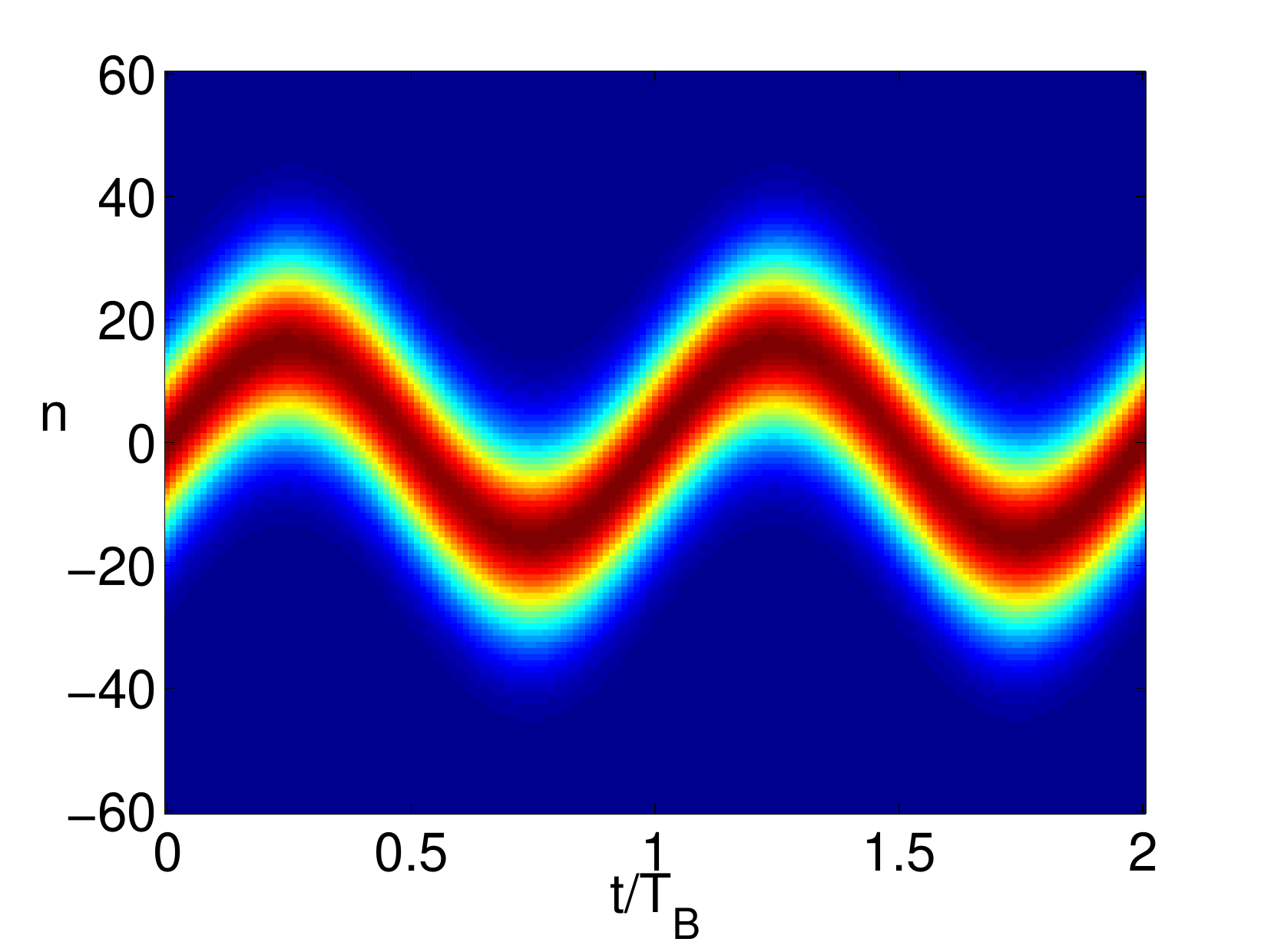}
\caption[ ]{Bloch oscillation: Breathing (left) and oscillatory mode (right) for an
initially narrow or extended distribution \cite{04bloch1d}.\label{f-bloch} }
\end{center}
\end{figure}
\noindent
The resulting wave function $\psi_n(t)$ is plotted as a color map in the left panel
of figure \ref{f-bloch}. This is a breathing mode, where the wave packet oscillates periodically
in the region $|n|<\frac{\Delta}{dF}\,|\sin \pi t/T_B|$ (see \cite{04bloch1d} for
more information). If a broad initial wave function is chosen, the dynamics
changes. For example  an initial Gaussian distribution, realized in the
program by replacing the line marked as \verb?initial state (localized)? by the lines
{\small
\begin{listing}[1000]{1001}
sig = 0.005; phi0 = pi/2;
psi = exp(-sig*n.^2+i*n*phi0)'; 
psi = psi/sqrt(sum(abs(psi).^2));
\end{listing}
}
\noindent
leads to an oscillating mode as shown in the right panel of figure \ref{f-bloch} known as the
Bloch oscillation.

\begin{figure}[htb]
\begin{center}
\includegraphics[width=80mm,clip]{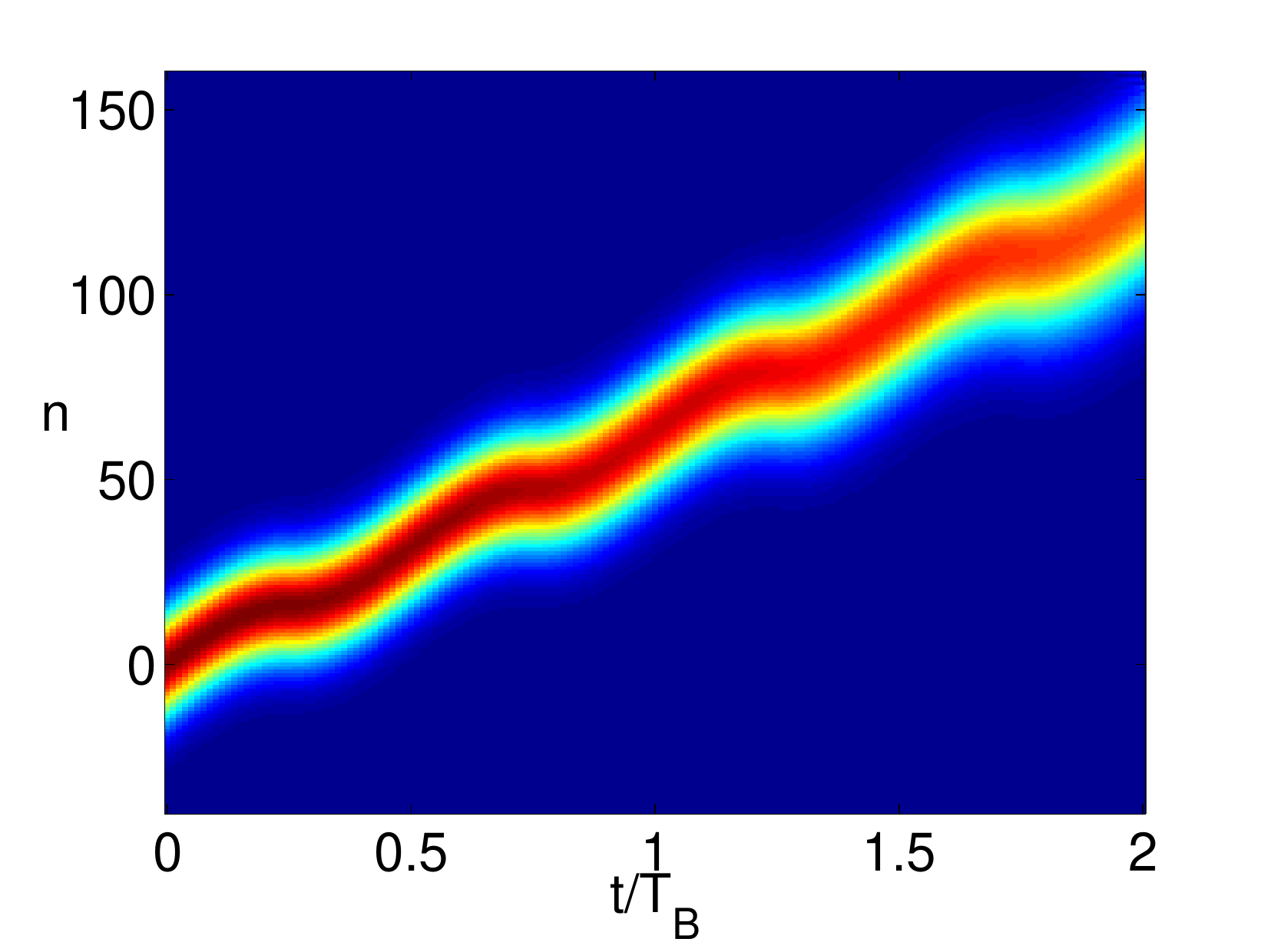}
\caption[ ]{Bloch oscillation: Directed almost dispersionless transport for a field 
flipped periodically with twice the Bloch frequency. \label{f-blochflip} }
\end{center}
\end{figure}

Most remarkably, such a Bloch oscillation leads
to an almost dispersionless directed transport during the first quarter of the Bloch period.
Then the motion continues in the opposite direction, which
suggests that we can prevent this backward motion by a flip of the field direction at this time. 
Continuing this process periodically, one can expect a directed dispersionless
transport. To study such a field-flip system numerically, we first replace 
the second line in the program above by \verb?nmax = 160;nmin = -40;? and the
Hamiltonian \verb?H?  by the two lines
{\small
\begin{listing}[1000]{1001} 
Hp =+d*F*diag(n,0)+(Delta/4)*(diag(m,-1)+diag(m,1));
Hm =-d*F*diag(n,0)+(Delta/4)*(diag(m,-1)+diag(m,1));
\end{listing}
}
\noindent
defining Hamiltonians with different signs of $F$. As an initial state, a Gaussian
is chosen and, for the time-propagation, the corresponding time-evolution 
operators are defined, which are then applied alternately during the subsequent Bloch periods:
{\small
\begin{listing}[1000]{1001}
Up = expm(-i*Hp*2*pi/d/F/J);
Um = expm(-i*Hm*2*pi/d/F/J);
nn = 0;
for nb = 1:2*N
  for nt = 1:J/4
    nn = nn+1;
    Psi(:,nn+1) = Up*Psi(:,nn);
  end
  for nt = 1:J/4
    nn = nn+1;
    Psi(:,nn+1) = Um*Psi(:,nn);
  end
end
\end{listing}
}
\noindent
The resulting dynamics shown in figure \ref{f-blochflip} agrees with
our conjecture. Most remarkably, however, one can show that the transport 
velocity is given by $v=\Delta d/\pi\hbar$, which is independent of the
force $F$ \cite{04bloch1d}.
%
\section{Angular momentum operators}
\label{s-ang}
In a similar manner one can construct the matrix representation
of the angular momentum operators $\hat J_x$, $\hat J_y$, $\hat J_z$ in the
basis of eigenstates $|j,m\rangle$ of the operator $\hat J_z$ with eigenvalues 
$m=-j,\,-j+1,\ldots,+j$, where $j\ge 0$ is the total angular momentum, which
is an even or odd multiple of $\case{1}{2}$. First we construct the ladder operators
$\hat J_+$ and $\hat J_-=\hat J_+^\dagger$ with 
$\hat J_+ |j,m\rangle =\sqrt{j(j+1)-m(m+1)}\,|j,m\!+\!1\rangle$
and the relations 
\begin{eqnarray}
\label{jxyz}
\hat J_x=\case{1}{2}\,\big(\hat J_-+\hat J_+\big)\ ,\quad
\hat J_y=\case{\rmi}{2}\,\big(\hat J_--\hat J_+\big)\ ,\quad
\hat J_z=\case{1}{2}\big[\hat J_+,\hat J_-\big]
\end{eqnarray}
coded in the following program lines for $j=2$:
{\small
\begin{listing}[1000]{1001}
j = 2; m = -j:j-1;
Jp = diag(sqrt(j*(j+1)-m.*(m+1)),1); Jm = Jp';
Jx = (Jm+Jp)/2;
Jy = -i*(Jm-Jp)/2;
Jz = (Jp*Jm-Jm*Jp)/2;
\end{listing}
}
\noindent
If desired, one can check here the angular momentum commutation
relations $[\hat J_x,\hat J_y]=\rmi \hat J_z$
 by means of \verb?Jx*Jy-Jy*Jx-i*Jz?, which should yield the zero matrix.
As an application, one can calculate the energy eigenvalues of the rigid body
Hamiltonian ($\hbar=1$)
\begin{eqnarray}
\label{rb_ham}
\hat H=\frac{\hat J_x^2}{2I_x}+\frac{\hat J_y^2}{2I_y}+\frac{\hat J_z^2}{2I_z}\,,
\end{eqnarray}
where the $I_x$, $I_y$ and $I_z$ are the principal moments of inertia.
For a symmetric top, the eigenvalues of the Hamiltonian are well known,
namely
\begin{eqnarray}
\label{E_rb_ham}
E_{jk}=\frac{j(j+1)}{2I_x}+\Big(\frac{1}{2I_z}-\frac{1}{2I_x}\Big)k^2
\ ,\quad k=-j,\,\ldots,j
\end{eqnarray}
for $I_x=I_y\ne I_z$. For an asymmetric top, however, the eigenvalues
for the special cases $j=1,\,2,\,3$ are given in \cite{Land77}, but no general formula
exists. This motivates, of course, a numerical approach, which is achieved by adding the program lines
{\small
\begin{listingcont}
Ix = 1/3; Iy = 1/2; Iz = 1;
H = Jx^2/2/Ix+Jy^2/2/Iy+Jz^2/2/Iz;
E = eig(H)'
\end{listingcont}
}
\noindent
in order to calculate the energy eigenvalues of an asymmetric top with
$I_x=1/3$, $I_y=1/2$ and $I_z=1$ for $j=2$. 
Note, however, that  for this system body-fixed angular momenta 
must be used with
commutation relations  $[\hat J_x,\hat J_y]=-\rmi \hat J_z$ (see, e.g.,
\cite{Land77,Ball06} for an explanation). This can be achieved by changing the
sign of the matrix representing $\hat J_y$. This subtlety does not affect the
eigenvalues of the Hamiltonian (\ref{rb_ham}) because it depends only of the squares
of the operators. The numerical eigenvalues\\
\verb?      E = 4.2679   4.5000   6.0000   7.5000   7.7321?\\
given by the program  agree with the formulas in \cite{Land77}:
\begin{eqnarray}
\label{rb-eig-2}
\fl &&E_{1}=\frac{2}{I_z}+\frac{1}{2I_x}+\frac{1}{2I_y}\ ,\quad 
E_{2}=\frac{2}{I_y}+\frac{1}{2I_z}+\frac{1}{2I_x}\ ,\quad
E_{3}=\frac{2}{I_x}+\frac{1}{2I_y}+\frac{1}{2I_z}\ ,\nonumber\\
\fl &&E_{4,5}=\frac{1}{I_x}+\frac{1}{I_y}+\frac{1}{I_z}
\pm\sqrt{\Big(\frac{1}{I_x}+\frac{1}{I_y}+\frac{1}{I_z}\Big)^2-
3 \Big(\frac{1}{I_xI_y}+\frac{1}{I_yI_z}+\frac{1}{I_zI_x}\Big)\,}\,.
\end{eqnarray}

\begin{figure}[htb]
\begin{center}
\includegraphics[width=80mm,clip]{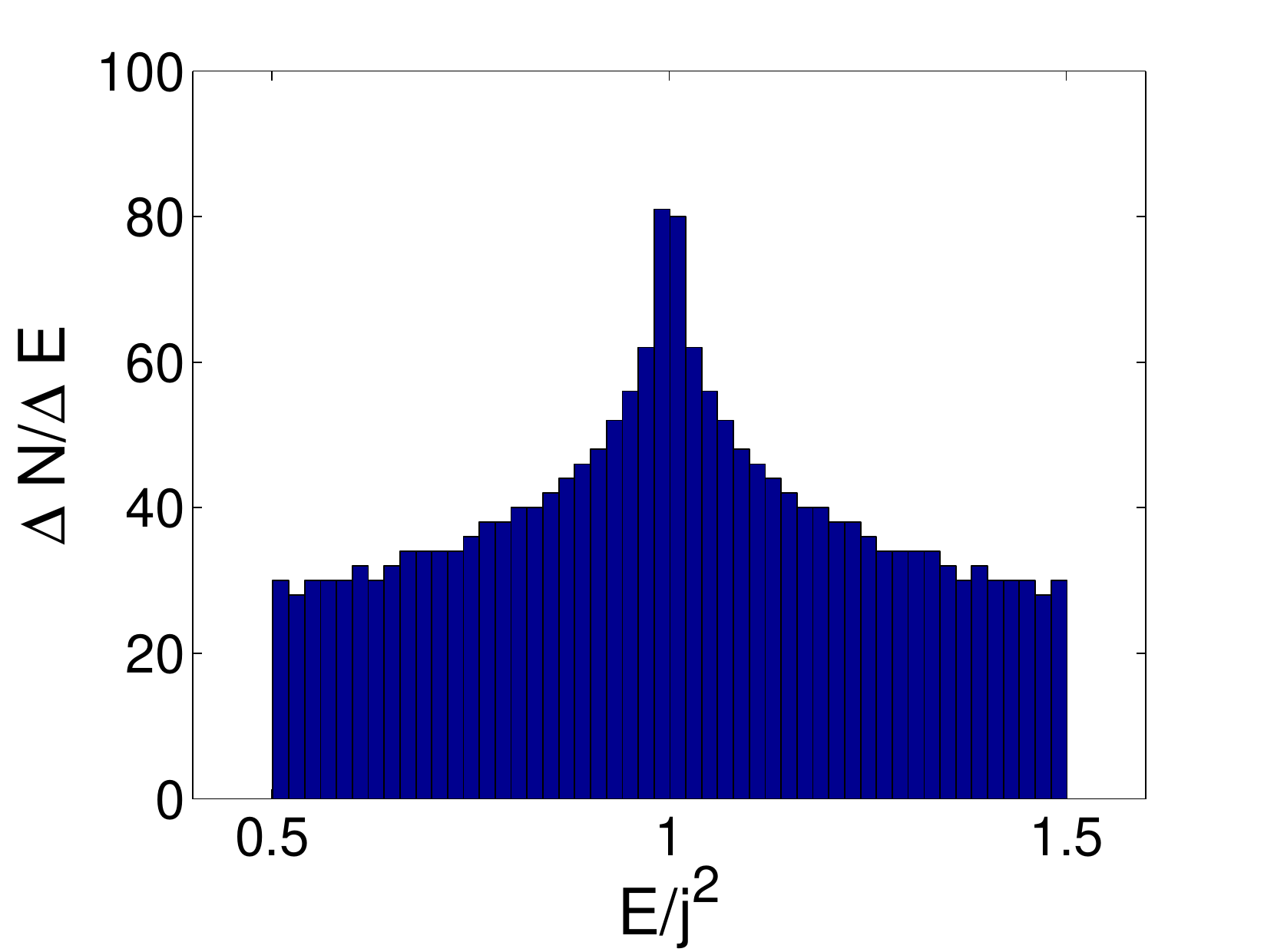}
\caption[ ]{Asymmetric top: Histogram of the  density of energy eigenvalues for rotational momentum 
$j=1000$ showing a pronounced  spike at the classical energy at the saddle point of the
energy surface. \label{f-asytop} }
\end{center}
\end{figure}

Such a small value as $j=2$ is, of course, not a challenge for a computational
treatment. The interested reader may try, for instance, the larger angular momentum
$j=1000$  and analyze the  distribution of the eigenvalues by plotting the state density via \verb?hist(E/j^2,50)?. The resulting distribution in figure \ref{f-asytop}
shows a clear restriction to an energy interval and a pronounced maximum.  This
structure can be understood by the observation that for large angular momentum
the system behaves almost classically. Here the classical dynamics is restricted
to a sphere with constant angular momentum $|\vec J\,|=j$ and the energy
function $H=\frac{J_x^2}{2I_x}+\frac{J_y^2}{2I_y}+\frac{J_z^2}{2I_z}$
possesses two minima, two maxima and two saddle points on the angular
momentum sphere. For the parameters used in the program, the energy 
at the minima is $E_{\rm min}/j^2=1/(2I_z)=0.5$, at the maxima
$E_{\rm max}/j^2=1/(2I_x)=1.5$, and at the saddle point $E_{\rm sad}/j^2=1/(2I_y)=1$.
This explains the restriction of the quantum energy eigenvalues to the
classically allowed region $E_{\rm min}<E<E_{\rm max}$.
Furthermore, the quantum state density at the extrema is approximately
equal to the period of the classical orbits divided by $2\pi$, which explains
the  spike  in the figure at the location of saddle point energy because the period at the saddle point is infinite.  Note that for $j \to \infty$ the quantum density 
also diverges at the saddle point energy.  
An additional application of the angular momentum operators can be found
in section \ref{s-many} below.
%
%
\section{Two-dimensional systems}
\label{s-two}
For a particle bound in a two-dimensional potential, the Hamiltonian
\begin{eqnarray}
\label{ham2}
\hat H=\frac{\hat p_1^2}{2m}+\frac{\hat p_2^2}{2m}+\hat V(\hat x_1,\hat x_2)\,,
\end{eqnarray}
can be conveniently expressed in the way described above by means
of the tensor product, i.e.~the Kronecker product \verb?kron? provided by 
\ML. Denoting the position operator for one degree of freedom as $\hat x$
and the corresponding identity operator as $\hat I$, the position operator
for the two particles are $\hat x_1=\hat x \otimes \hat I$  and  
$\hat x_2=\hat I \otimes \hat x$. The same expressions appear for the
momentum operators.  

As an example the program described below computes the lowest energy
eigenvalues for the Pullen-Edmonds potential \cite{Pull81}
\begin{eqnarray}
\label{VPE}
V(x_1,x_2)=\case{1}{2}x_1^2+\case{1}{2}x_2^2 +\alpha x_1^2x_2^2
\end{eqnarray}
for $\alpha=0.5$. This potential has been
employed in a number of studies related to quantum chaos and also found
applications to various molecular systems.
The symmetry group of the Hamiltonian is $C_{4v}$
and the eigenstates can be classified by the irreducible representation
${\cal A}_{1,2}$,  ${\cal B}_{1,2}$ and ${\cal E}$
 \cite{Pull81}. The  wave functions with symmetry ${\cal A}_1$ or ${\cal B}_2$
 are symmetric if $x_1$ and $x_2$ are interchanged,  those with
 ${\cal A}_2$ or ${\cal B}_1$ symmetry antisymmetric. If, on the other hand,
 the $x_j$ are changed to $-x_j$,  the ${\cal A}_1$ or ${\cal B}_1$ 
 wave functions are unaffected  and the 
 ${\cal A}_2$ or ${\cal B}_2$ wave functions change sign. In a harmonic oscillator
 expansion 
\begin{eqnarray}
\label{psiexp}
\varphi(x_1,x_2)=\sum_{n,m=0}^\infty C_{nm}\varphi_n(x_1)\varphi_m(x_2)
\end{eqnarray}
the coefficients satisfy $C_{nm}=C_{mn}$ for ${\cal A}_1$ or ${\cal B}_2$
and  $C_{nm}=-C_{mn}$ for ${\cal A}_2$ or ${\cal B}_1$. 
In addition only coefficients with even indices appear for  ${\cal A}_1$ or ${\cal B}_1$
and only coefficients with odd indices for  for  ${\cal A}_2$ or ${\cal B}_2$.
The states with ${\cal E}$ symmetry are twofold degenerate.
The following program generates the one-dimensional operators as $N\times N$ matrices
and constructs the Hamiltonian as $N^2\times N^2$ matrices using the tensor 
product as described above. The first $n_{\rm out}$  eigenvalues  are 
displayed on the screen and the wave function of state number $n_{\rm plot}$ 
is finally plotted.
{\small
\begin{listing}[1000]{1001}
N = 10; nout = 6; nplot = 4; 
m = sqrt(1:N-1); md = diag(m,-1);
x = 1/sqrt(2)*(md + md');
p = i/sqrt(2)*(md - md');
I = eye(N);
x1 = kron(x,I); x2 = kron(I,x); p1 = kron(p,I); p2 = kron(I,p);
alpha = 0.5;
H = p1^2/2+p2^2/2+x1^2/2+x2^2/2+alpha*x1^2*x2^2;
[C,Eig] = eig(H);
format short
E = diag(Eig)(1:nout)'
\end{listing}
}
\noindent
This resulting eigenvalues\\
\verb?     1.0980   2.2634   2.2634   3.2791   3.5157   3.7214?\\
agree very well with those given in \cite{Amor09} for the ground state
energies
with ${\cal A}_1$, ${\cal E}$,  ${\cal B}_1$ and ${\cal B}_2$  symmetry:\\[2mm]
\verb?     1.0980   2.2634   3.2789   3.7223?\\[2mm]
In order to identify the symmetry of the states calculated numerically, 
the matrix of the expansion coefficients is displayed by means of
{\small
\begin{listingcont}
format bank
Cplot = reshape(C(:,nplot),N,N)
\end{listingcont}
}
\noindent
for the state $N$ with energy $3.27914$, whose wave function is subsequently 
plotted. One observes indeed that for this state
the matrix is antisymmetric and that all entries with odd indices vanish as
required for a  symmetry ${\cal B}_1$. This symmetry is, of course, also
visible if the wave function $\varphi(x_1,x_2)$ is explicitly calculated and
plotted as a color map. The following program lines first compute the
one-dimensional harmonic oscillator wave functions $\varphi_n(x)$ iteratively
which are then stored in the matrix \verb?hermval?
as described in  \cite{02computing}. Finally the expansion (\ref{psiexp})
is carried out as the scalar product \verb?psi = hermval*Cplot*hermval'? and plotted:
{\small
\begin{listingcont}
xx = -4:0.05:4;
Nx = length(xx);
hermval = zeros(Nx,N);
h0 = [1];
hermval(:,1) = polyval(h0,xx).*exp(-0.5*xx.^2);
h1 = [sqrt(2) 0];
hermval(:,2) = polyval(h1,xx).*exp(-0.5*xx.^2);
v1 = [1 0]; v0 = [0 0 1];
for n = 2:N-1  % recursion
  h2 = sqrt(2/n)*conv(h1,v1)-sqrt(1-1/n)*conv(h0,v0);
  h0 = h1; h1 = h2;
  hermval(:,n) = polyval(h2,xx).*exp(-0.5*xx.^2);
end 
psi = hermval*Cplot*hermval';
imagesc(xx,xx,psi); axis square;
set(gca,'Fontsize',20)
xlabel('x','FontSize',24); ylabel('y','Rotation',0,'FontSize',24);
\end{listingcont}
}
\noindent
For more pictures of Pullen-Edmonds eigenfunctions see, e.g., \cite{Jose14}.
\begin{figure}[htb]
\begin{center}
\includegraphics[width=80mm,clip]{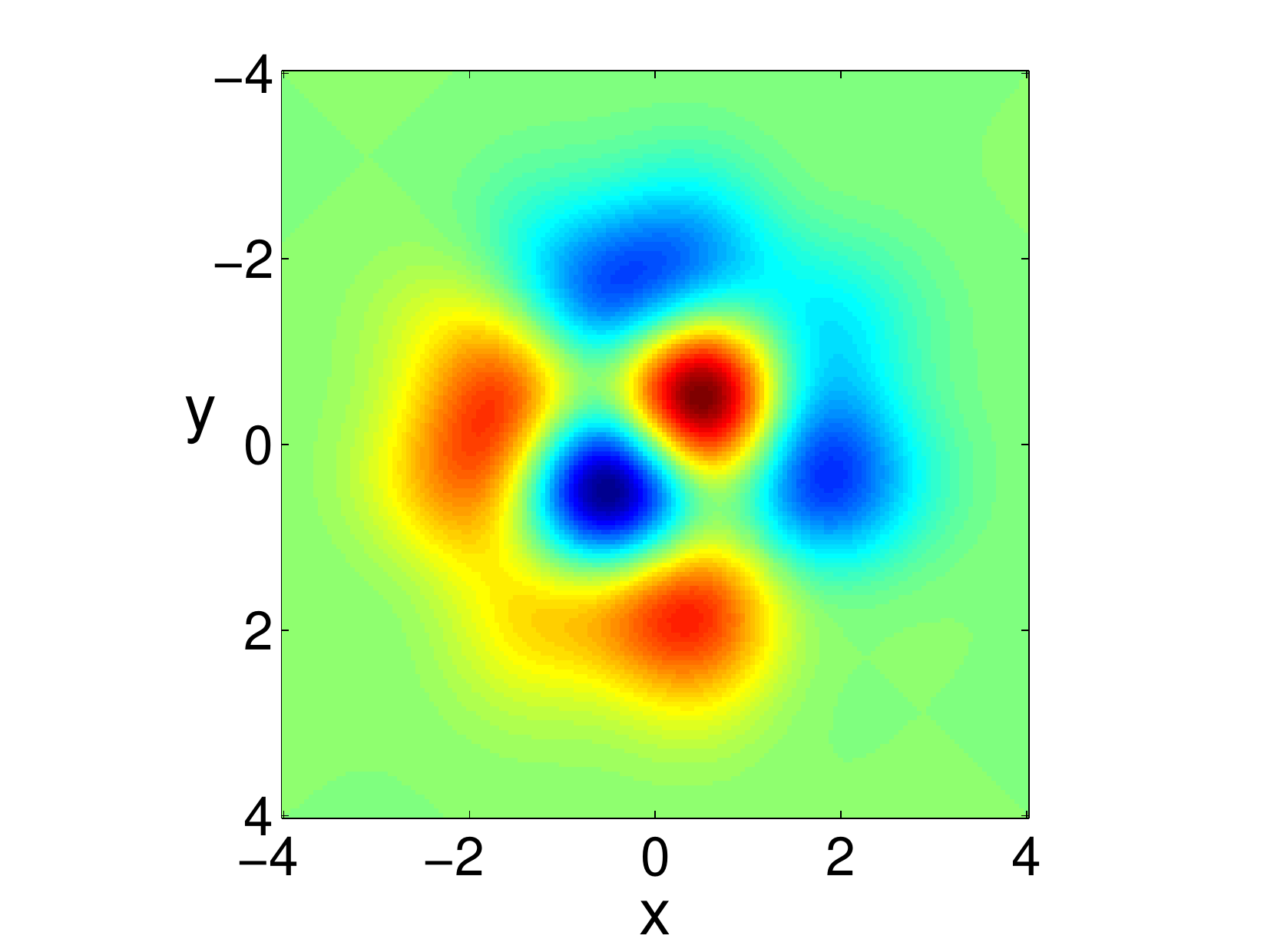}
\caption[ ]{Pullen-Edmonds potential: Color map of the wave function of the lowest
eigenstate with symmetry ${\cal B}_1$. \label{f-eigen2} }
\end{center}
\end{figure}
%
\section{Many-particle systems}
\label{s-many}
A prominent example of a  many-particle quantum system  is the $N$-particle 
Bose-Hubbard system \cite{Pita03}, which is one of the
basic models studied in theoretical investigations of the dynamics of many particles on a
lattice. Numbering the lattice sites by $n$, the operators $\hat a_n$ and $\hat a_n^\dagger$ 
with $[\hat a_n,\hat a_n^\dagger]=1$ describe annihilation and creation of a particle
at site $n$ and $\hat a_n^\dagger\hat a_n$ is the particle number operator
at site $n$. These operators for different sites commute. Then the hermitian operators
$\hat a_{n+1}^\dagger \hat a_{n}$ describe the
destruction of a particle at site $n$ and the creation at site $n+1$, and 
$\hat a_n^\dagger \hat a_{n+1}+\hat a_{n+1}^\dagger \hat a_{n}$
the hopping of a particle between these sites. If the particles interact with each other
the interaction energy is proportional to the product of the particle number operators.
In many cases this interaction is short ranged, so that only the interaction of
particles on the same site must be taken into account. Let us confine ourselves here
to the simple case of a two-site system, the Bose-Hubbard dimer, with Hamiltonian
\begin{eqnarray}
\label{Hbhdimer}
\hat H=\epsilon\big(\hat a_1^\dagger \hat a_1-  \hat a_2^\dagger \hat a_2 \big)
+v\big(\hat a_1^\dagger \hat a_2+\hat a_2^\dagger \hat a_1\big)
+c\big(\hat a_1^\dagger \hat a_1-\hat a_2^\dagger \hat a_2\big)^2\,,
\end{eqnarray}
where $\pm \epsilon$ are the site energies, $v$ the hopping and $c$ the
interaction strength. The Hamiltonian commutes with the
particle number operator $\hat N=\hat a_1^\dagger \hat a_1+\hat a_2^\dagger \hat a_2$,
i.e.~the particle number $N$ is conserved. 

It is of interest to realize that the Bose-Hubbard dimer also appears naturally for 
a collection of $N$ bosonic atoms in a double-well potential, which is deep enough so
that only the lowest state in each well is populated. In this two-mode 
approximation the system can be described by the Hamiltonian (\ref{Hbhdimer}).

The following \ML program calculates the eigenvalues of the Hamiltonian (\ref{Hbhdimer}).
First the creation and annihilation operators at the two sites are constructed by
means of the \verb?kron? product as well as  the Hamiltonian, whose eigenvalues and
eigenstates are then calculated for  $N=24$, i.e.~a matrix dimension of $N_p=N+1=25$ for the single particle operators  $\hat a$:
{\small
\begin{listing}[1000]{1001}
N = 24; Np = N+1; Nout = 10; epsilon = 1; v = 1; c = 1;
a = diag(sqrt(1:N),1); ad = a'; I = eye(Np);          
a1 = kron(a,I); ad1 = a1';        
a2 = kron(I,a); ad2 = a2';         
H = epsilon*(ad1*a1-ad2*a2)+v*(ad1*a2+ad2*a1)+c/2*(ad1*a1-ad2*a2)^2; 
[C,E] = eig(H); E = diag(E);
\end{listing}
}
\noindent
One should be aware of the fact that the many-particle operators  are represented
by $N_p^2\times N_p^2$ matrices and therefore one obtains $N_p^2=625$
energy eigenvalues, however not all of them are fully converged because of the
restricted basis set $|n_1,n_2\rangle,\ n_{1,2}=0,\,1,\,\ldots ,\,N$. All eigenstates
populating only this restricted basis set are accurately represented, i.e.~those with
particle numbers up to $N$, a number of $N_p(N_p+1)/2$ states.
To illustrate this, the following program lines compute for all
calculated eigenstates the expectation values $\langle \hat N\rangle$ of the number 
operator, which agree with the exact particle number for the eigenstates because
$\hat H$ and $\hat N$ commute. These values are subsequently ordered. 
There are $N'+1$ eigenstates for each value of $N'=\langle \hat N\rangle\le N$,
and therefore a total number of $N_p(N_p+1)/2$, which are converged for the
chosen basis size $N_p$. Finally the converged energy eigenvalues are plotted
as a function of the particle number $N'$. 
{\small
\begin{listingcont}
Nav = diag(C'*(ad1*a1+ad2*a2)*C);
[Nav,index] = sort(Nav);
E = E(index);
plot(Nav(1:Np*(Np+1)/2),E(1:Np*(Np+1)/2),'b*')
set(gca,'ydir','normal','FontSize',20); axis([0 25 -50 350])
xlabel('<N>','FontSize',20); ylabel('E_n','rotation',0,'FontSize',20)
\end{listingcont}
}
\begin{figure}[htb]
\begin{center}
\includegraphics[width=80mm,clip]{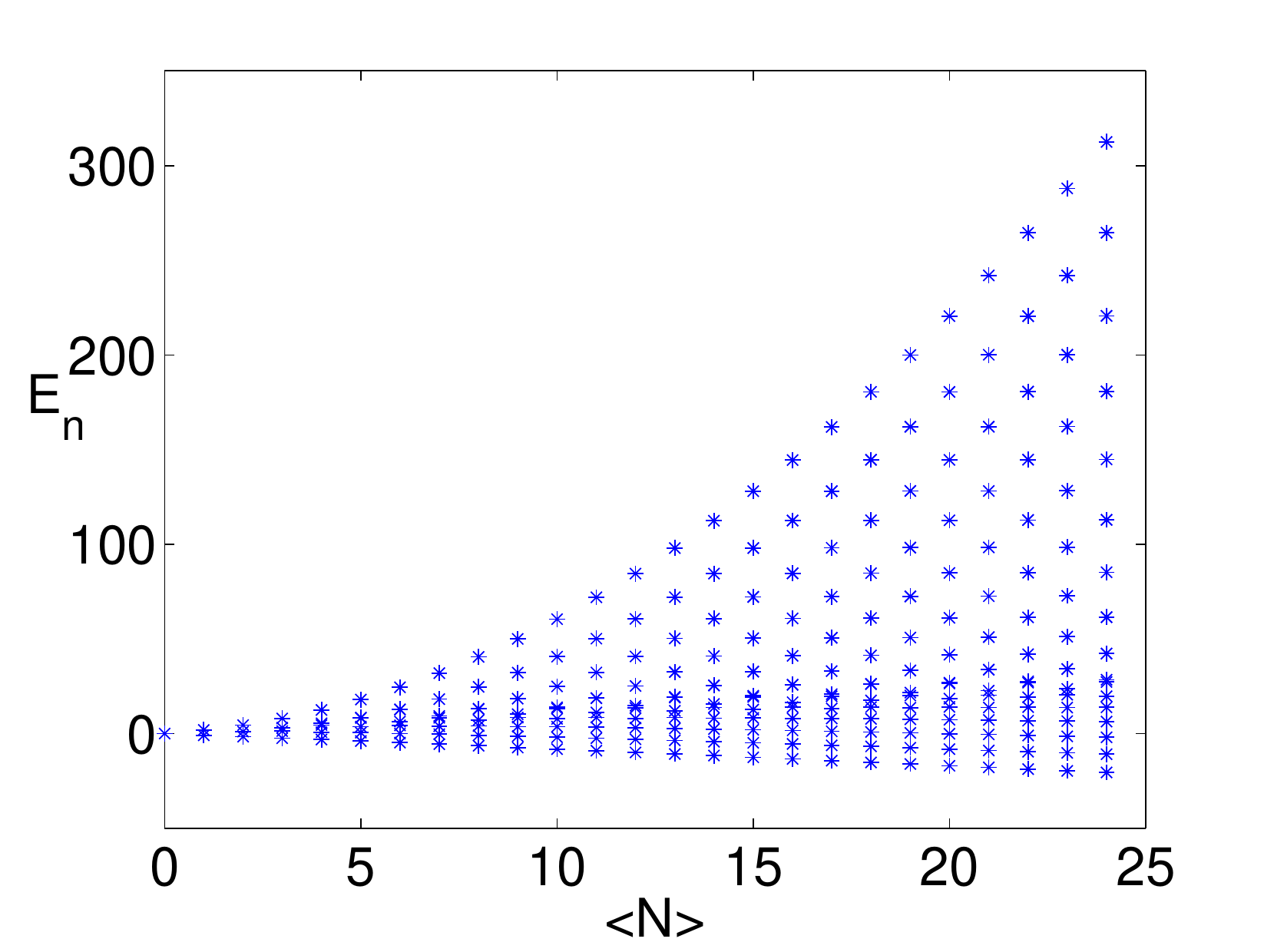}
\caption[ ]{Bose-Hubbard dimer: Energy eigenvalues as a function of the
particle number $N$. \label{f-bhdimer} }
\end{center}
\end{figure}

In many applications, one is only interested in the eigenvalues and eigenstates
of the Hamiltonian for a prescribed value $N$ of the particle number.
Then one can make use of the following trick: An extra term 
$-\lambda \big(\hat{a}_1^\dagger\hat{a}_1+\hat{a}_2^\dagger\hat{a}_2 -N\big)$ 
($\lambda \gg \epsilon, v$) is added to the Hamiltonian which does not affect the states 
with the given particle number $N$ but which shifts the states with different particle 
numbers to high energies. The desired eigenvalues can then be obtained by extracting 
the low-energy part of the spectrum of the resulting effective Hamiltonian as implemented 
in the following program:
{\small
\begin{listing}[1000]{1001}
N = 6; Np = N+1;
epsilon = 1; v = 1; c = 1;
a = diag(sqrt(1:Np-1),1); ad = a'; I = eye(Np);          
a1 = kron(a,I); ad1 = a1';        
a2 = kron(I,a); ad2 = a2';         
H = epsilon*(ad1*a1-ad2*a2)+v*(ad1*a2+ad2*a1)+c/2*(ad1*a1-ad2*a2)^2; 
lambda = 10000;
H1 = H-lambda.*(ad1*a1+ad2*a2-N*kron(I,I)); 
E1 = eig(H1);
E2 = E1.*(abs(E1)<10*epsilon.*N);
E = E2(find(E2))'    
\end{listing}
}
\noindent
Calculated are the seven eigenvalues for $N=6$ with the result\\
\verb? E = -4.792349  0.078611 4.347781 6.348595 12.731351 12.786011 24.500000?\\

A more sophisticated alternative method leading to the same results for the eigenvalues consists in computing the Hamiltonian $\hat H_N$ for a fixed particle number $N$ by projecting the original Hamiltonian $\hat H$ on the corresponding eigenstates of the number operator $\hat{a}_1^\dagger\hat{a}_1+\hat{a}_2^\dagger\hat{a}_2$. Since the latter is a represented by a diagonal matrix, the implementation is rather straightforward:
{\small
\begin{listing}[1000]{1001}
N_op = ad1*a1+ad2*a2;
iN=find(abs(diag(N_op)-N)<10^-6);
N_p=N_op(:,iN)./N;
H_N = N_p'*H*N_p;
E = eig(H_N)
\end{listing}
}
\noindent

For the Bose-Hubbard dimer, there is, however, a much more convenient way to describe the 
system for a fixed particle number $N$. The Jordan-Schwinger representation
\begin{eqnarray}
\hat J_x&=&(\hat a_1^\dagger\hat a_2+\hat a_2^\dagger\hat a_1)/2,\nonumber\\
\hat J_y&=&(\hat a_1^\dagger\hat a_2-\hat a_2^\dagger\hat a_1)/2\rmi, \\
\hat J_z&=&(\hat a_1^\dagger\hat a_1-\hat a_2^\dagger\hat a_2)/2\nonumber
\end{eqnarray}
transforms the system to angular momentum operators. The Hamiltonian
(\ref{Hbhdimer}) then takes the form
\begin{eqnarray}
\label{HbhdimerJ}
\hat H=2\epsilon\hat J_z+2v \hat J_x+2c\hat J_z^2
\end{eqnarray}
and the total angular momentum is $\hat J=\hat N/2$.  
Using the matrix representation of angular momentum operators
discussed in section \ref{s-ang}, this Hamiltonian can be easily coded
in the following \ML program:
{\small
\begin{listing}[1000]{1001}
N = 6; j = N/2; m = -j:j-1;
Jp = diag(sqrt(j*(j+1)-m.*(m+1)),1);
Jm = Jp';
Jx = (Jm+Jp)/2;
Jy = i*(Jm-Jp)/2;
Jz = (Jp*Jm-Jm*Jp)/2;
epsilon = 1; v = 1; c = 1; 
H = 2*epsilon*Jz+2*v*Jx+2*c*Jz^2;
E = eig(H)'
\end{listing}
}
\noindent
The results agree precisely with those given above for $N=6$.

In the case of high matrix dimensions time and memory can be saved by using sparse matrices.
In \ML this can be implemented straightforwardly, e.g.~the sparse matrix representation of the 
annihilation operator and the identity in an $N+1$-dimensional space are given by 
\texttt{a = sparse(diag(sqrt(1:N),1))} and  \texttt{I = speye(N+1)} respectively. This automatically 
leads to a sparse matrix representation of the Hamiltonian ${\hat H}$.
This already becomes relevant if the Bose-Hubbard dimer is extended by an additional site.
Such a Bose-Hubbard trimer, as described by the Hamiltonian
\begin{eqnarray}
\label{Hbhtrimer}
\hat H=\sum_{j=1}^3 \Big[-\frac{K}{2}\Big( \re^{\ri \Phi/3}\,\hat a^\dagger_{j+1} \hat a_j 
+ \re^{-\ri \Phi/3}\,\hat a^\dagger_{j} \hat a_{j+1} \Big)+ \frac{U}{2}\, 
\hat a_j^\dagger \hat a_j^\dagger \hat a_j \hat a_j \Big]
\end{eqnarray}
was considered in \cite{Arwa14}. Here we identify site number 4 with site number 1, i.e.~the 
three sites form a triangular structure which constitutes a minimal model for a superfluid circuit 
of a Bose-Einstein condensate. The additional phase factors in the hopping term describe the 
influence of an applied magnetic flux $\Phi$ (in scaled dimensionless units) or alternatively a 
rotation of the system with a frequency $\Omega \propto \Phi$ (see  \cite{Arwa14} for more details).

The following program computes the eigenvalues $E_n$, eigenstates $|n \rangle$ and the 
expectation values of the current 
$J_n=\langle n|\hat J|n\rangle =\langle n|\frac{\partial \hat H}{\partial \Phi}|n\rangle$ 
and plots the results:
{\small
\begin{listing}[1000]{1001}
ND = 3*10+3; N = ND-1
K = 1; Phi = 0.8*pi; u = 0.5
U = K*u/N
u_qm = 3*u/N.^2
a = sparse(diag(sqrt(1:ND-1),1)); ad = a'; I=speye(ND);   
a1 = kron(kron(a,I),I); ad1 = a1';        
a2 = kron(kron(I,a),I); ad2 = a2';         
a3 = kron(kron(I,I),a); ad3 = a3';          
N_op = ad1*a1+ad2*a2+ad3*a3;
iN = find(abs(diag(N_op)-N)<10^-6);
N_p = N_op(:,iN)./N;
H = -K/2*(exp(i*Phi/3)*(ad2*a1+ad3*a2+ad1*a3) ...
    +exp(-i*Phi/3)*(ad1*a2+ad3*a1+ad2*a3)) ...
    +U/2*(ad1^2*a1^2+ad2^2*a2^2+ad3^2*a3^2); 
H_N = N_p'*H*N_p;
[C,E] = eig(full(H_N)); E = diag(E)/u;
dH_dPhi = -K/2*(-i/3*exp(i*Phi/3)*(ad2*a1+ad3*a2+ad1*a3) ...
          +i/3*exp(-i*Phi/3)*(ad1*a2+ad3*a1+ad2*a3));
dH_dPhi_N = N_p'*dH_dPhi*N_p;
Jav = real(diag(C'*dH_dPhi_N*C)); 
plot(Jav,E,'o','markerfacecolor','r','markeredgecolor','r')
set(gca,'ydir','normal','FontSize',20); 
xlabel('<J>','FontSize',20); ylabel('E_n/u','rotation',90,'FontSize',20)
axis square
\end{listing}
}
\noindent
(depending on the \ML version used it may be required to replace 
\verb?eig(H_N)?  by \verb?eig(full(H_N))?.)

\begin{figure}[tb]
\begin{center}
\includegraphics[width=50mm,clip]{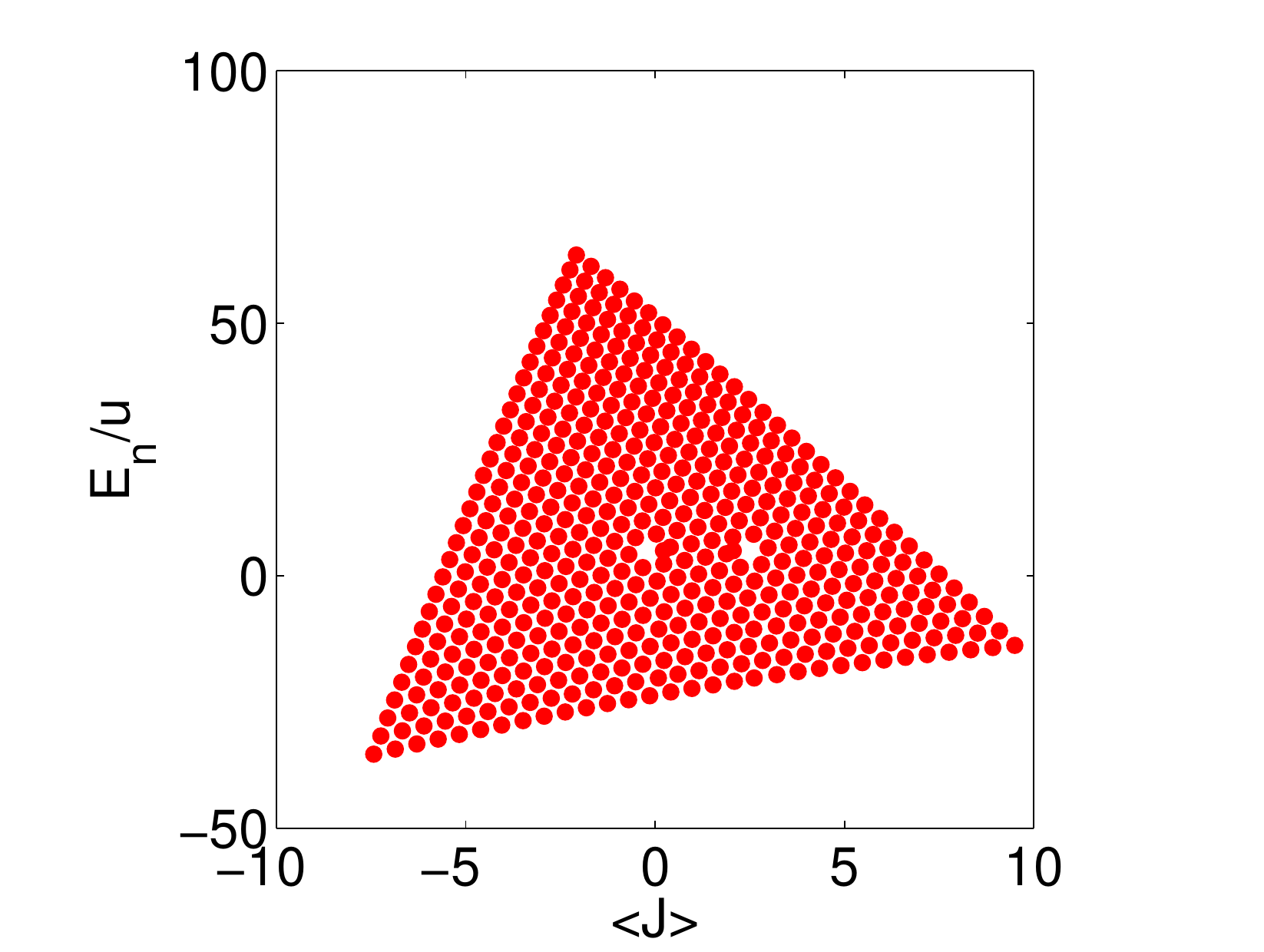}
\includegraphics[width=50mm,clip]{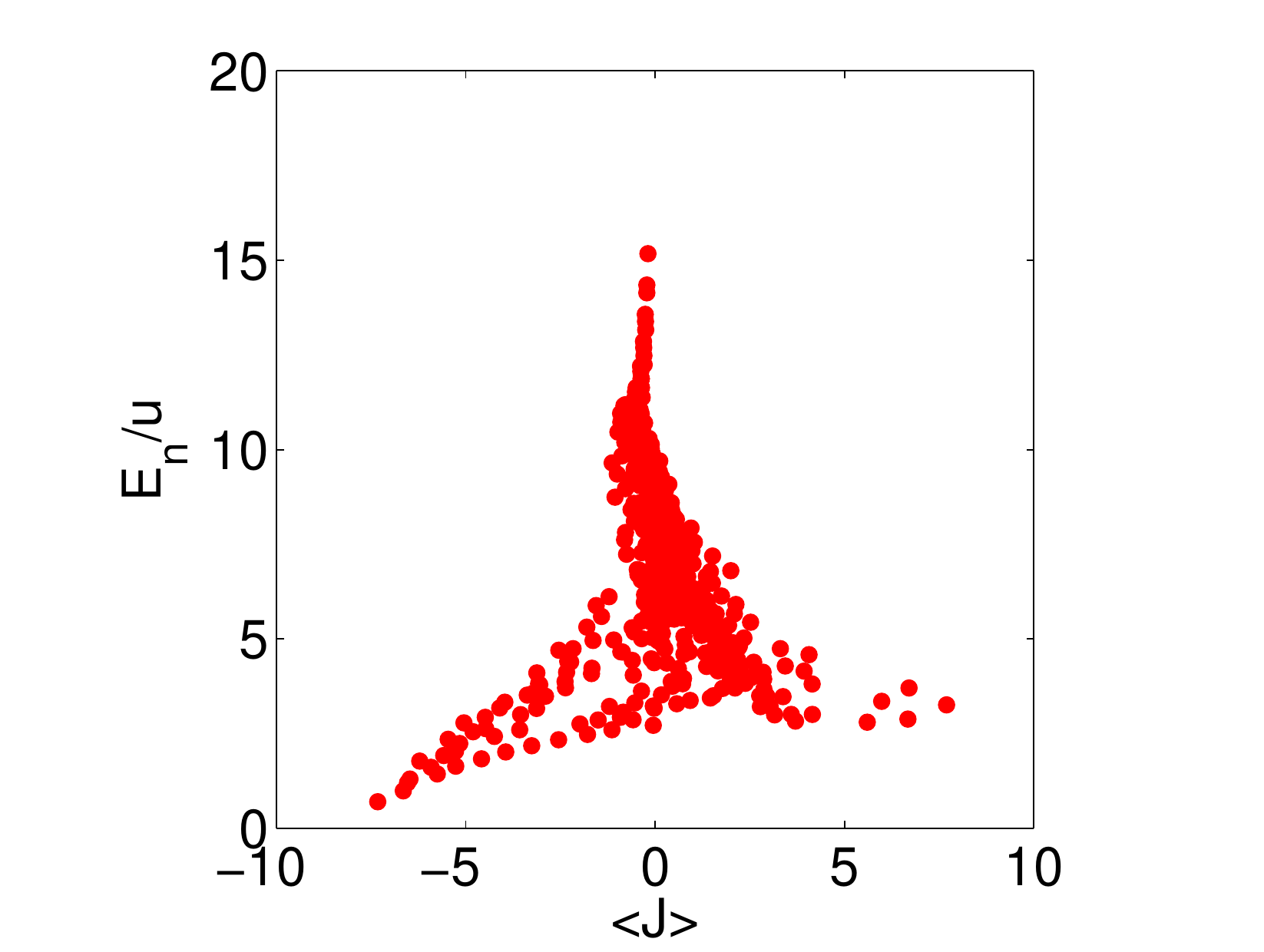}
\includegraphics[width=50mm,clip]{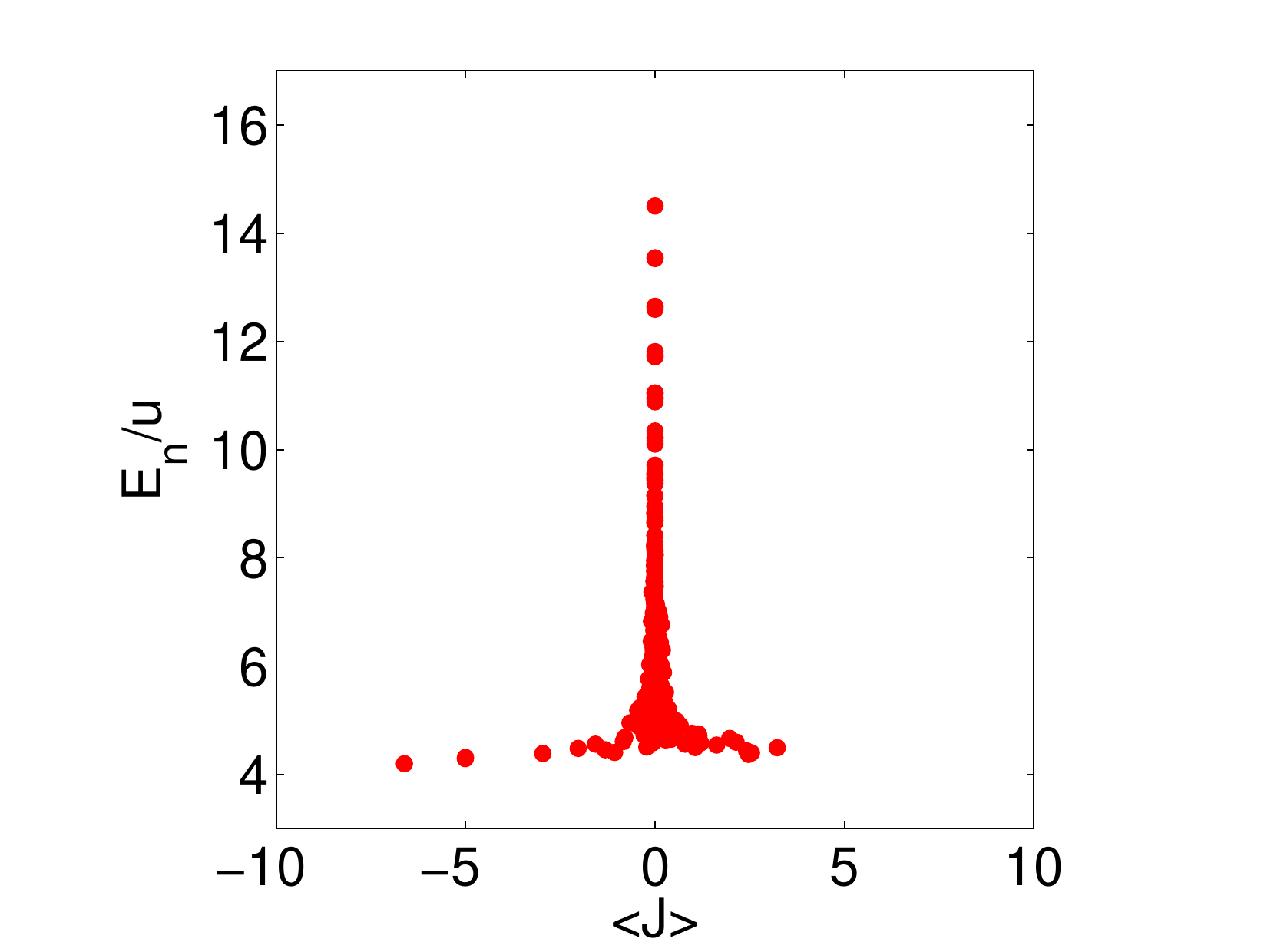}
\includegraphics[width=50mm,clip]{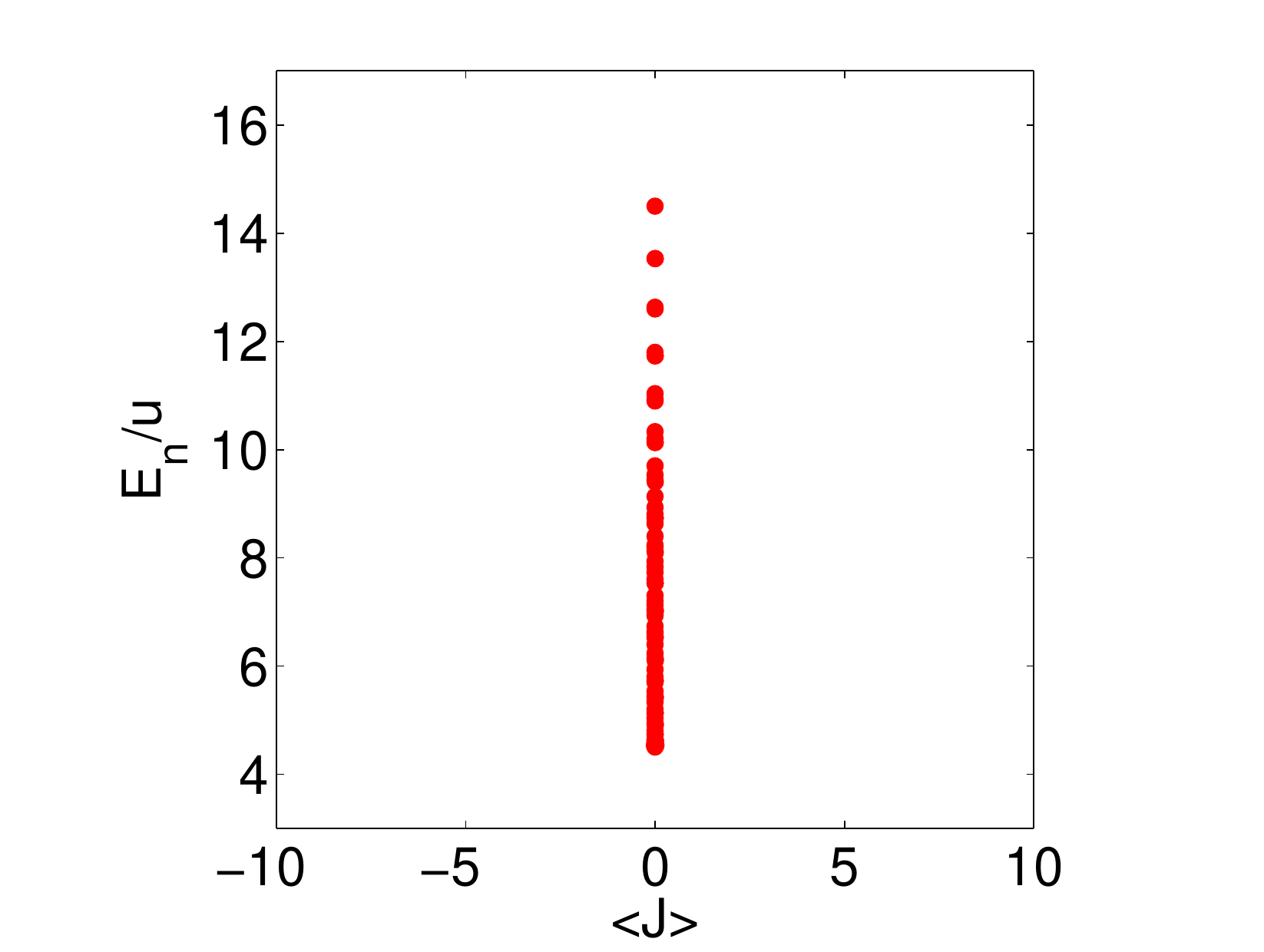}
\includegraphics[width=50mm,clip]{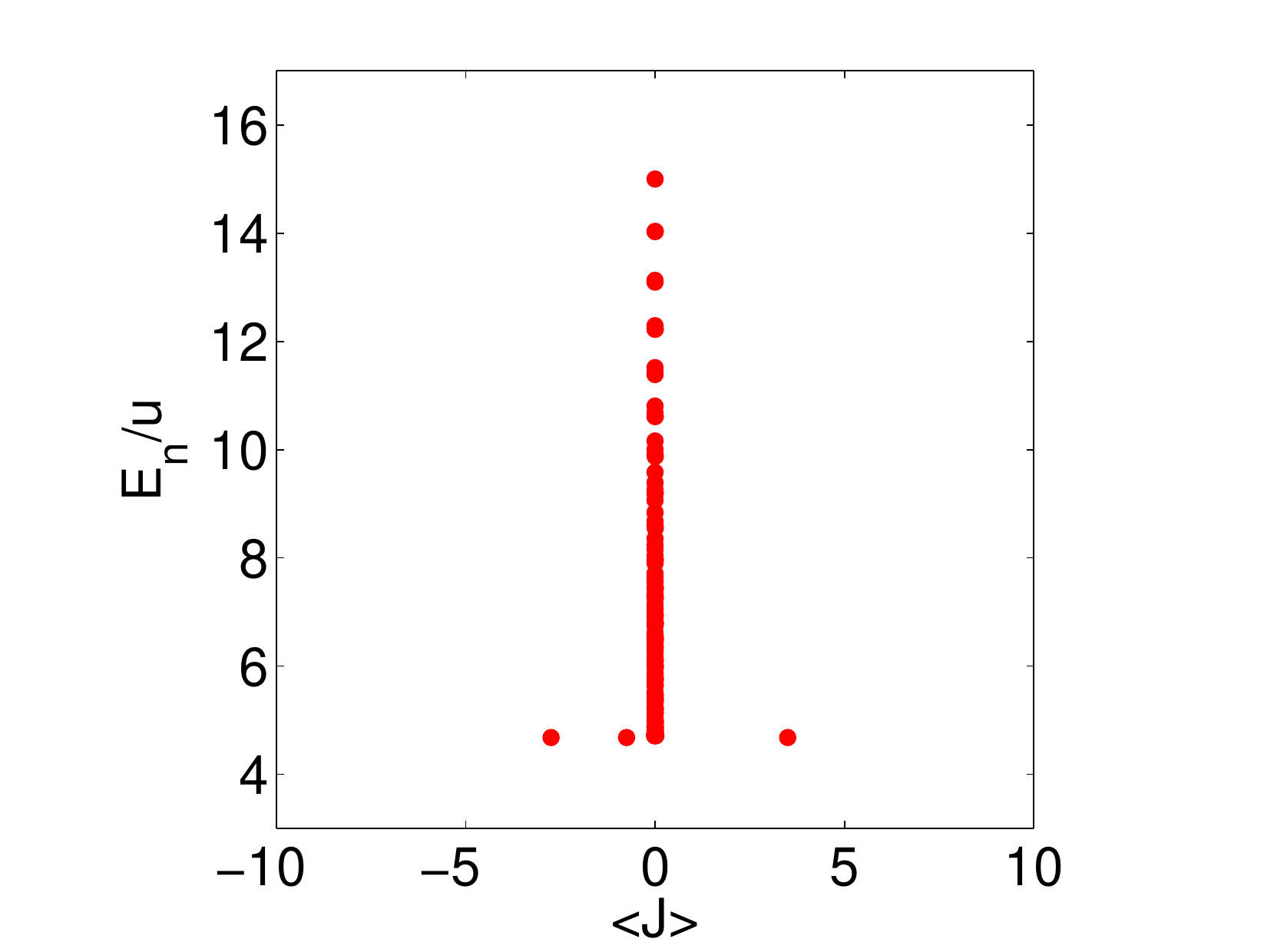}
\includegraphics[width=50mm,clip]{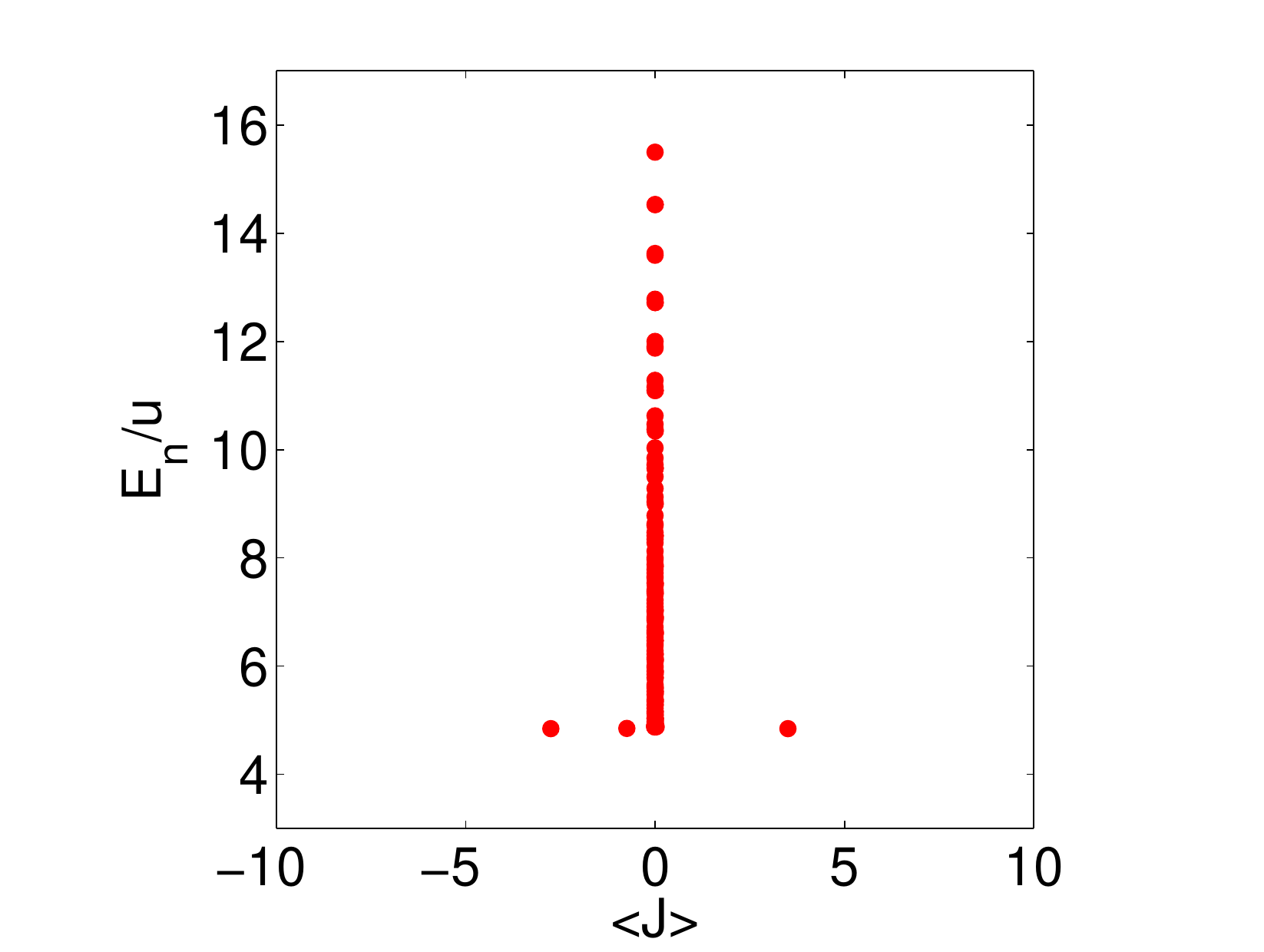}
\caption[ ]{Trimer:  \label{f-trimer} 
Current $\langle J\rangle$ and eigenenergies $E_n/u$ of the Bose-Hubbard trimer for $K=1$, 
$\Phi=0.8 \pi$. First four panels:
$N=30$ particles and interaction strength $u=UN/K=0.5,\,5,\,50,\,50000$ 
respectively. Last two panels: $N=31,\,N=32$ and $u=50000$.
}
\end{center}
\end{figure}

Figure \ref{f-trimer} shows the distribution of the eigenstates with respect to the eigenenergies 
and the currents. For noninteracting particles, $u=UN/K=0$, the current $\hat J$ commutes 
with the Hamiltonian and each eigenstate carries a quantized current, forming a triangle in the 
energy-current plane as explained in  \cite{Arwa14}.
For an integer filling $N=30=10\cdot3$ of the trimer, the eigenstates are more and more 
redistributed along the line indicating zero current with increasing interaction $u$. In the high 
interaction limit, even the ground state becomes a zero current and thus an insulator state. 
This is a signature of a so-called superfluid to Mott insulator transition, illustrated by the first four 
panels. In contrast, for a non-integer filling per site
the ground state always carries a current even in the strong interaction limit as shown for  $N=31$ and
 $N=32$ in the last two panels. More about the rich behavior of this three site Bose-Hubbard system
 can be found in \cite{Arwa14}.
%
\section{The Lindblad master equation for an open quantum system}
\label{s-lindblad}
%
Open quantum systems are relevant in various fields including electronic transport in 
semiconductors and nanostructures \cite{DiVe08} and cavity QED \cite{Scul97}.
Another example of an open system is given by Bose-Einstein condensates in optical 
lattices which are subject to a decay process due to coupling to an environment, a 
setup studied recently experimentally (see e.g.~\cite{Wuer09}) and theoretically using 
different approaches \cite{10nlret,Lode09,12decoh,08nhbh_s,Witt08,Rape13}. 
For weak decay such systems can be described by means of a Born-Markov approximation 
leading to a Lindblad master equation for the system's density matrix \cite{Breu02}. For the 
Bose-Hubbard dimer of section \ref{s-many} the Lindblad master equation can be written as 
(see e.~g.~\cite{Witt08})
\begin{equation}
   \dot{{\hat \rho}}=-\rmi \big[{\hat H},{\hat \rho}\,\big]-\frac{\gamma }{2}\,
   \Big( {\hat a}_2^\dagger {\hat a}_2{\hat \rho}+ {\hat \rho}{\hat a}_2^\dagger 
   {\hat a}_2-2 {\hat a}_2 {\hat \rho}{\hat a}_2^\dagger \Big)
\label{Lindblad}
\end{equation}
where the second term models a particle decay from site $2$ with rate $\gamma$ and the 
first term corresponds to the hermitian part of the time evolution with the two site Bose-Hubbard 
Hamiltonian ${\hat H}$ as given in equation (\ref{Hbhdimer}). In \cite{Rape13} it was shown that 
this equation accurately describes tunneling decay from the dimer into a weakly coupled optical lattice.

In the following we consider the situation where initially $N$ particles are in the second well of 
the Bose-Hubbard dimer whereas the first well is empty. The corresponding initial wavefunction and 
density matrix at time $t=0$ are then given by $|\psi_0\rangle=|0\rangle \otimes |N\rangle $  and 
$\hat{\rho}(0)=|\psi_0\rangle\langle \psi_0|$ respectively. 
Using the matrix representations introduced in the previous sections the following \ML code
propagates the initial density matrix according to equation (\ref{Lindblad}) by means of a 
predictor corrector integrator \cite{Pres07}. The expectation values of the time-dependent site occupations 
are then obtained via $n_j(t)= \mathrm{trace}\,\big({\hat \rho}(t)\,{\hat a}_j^\dagger {\hat a}_j\big)$.

\begin{figure}[htb]
\begin{center}
\includegraphics[width=0.5\textwidth,clip]{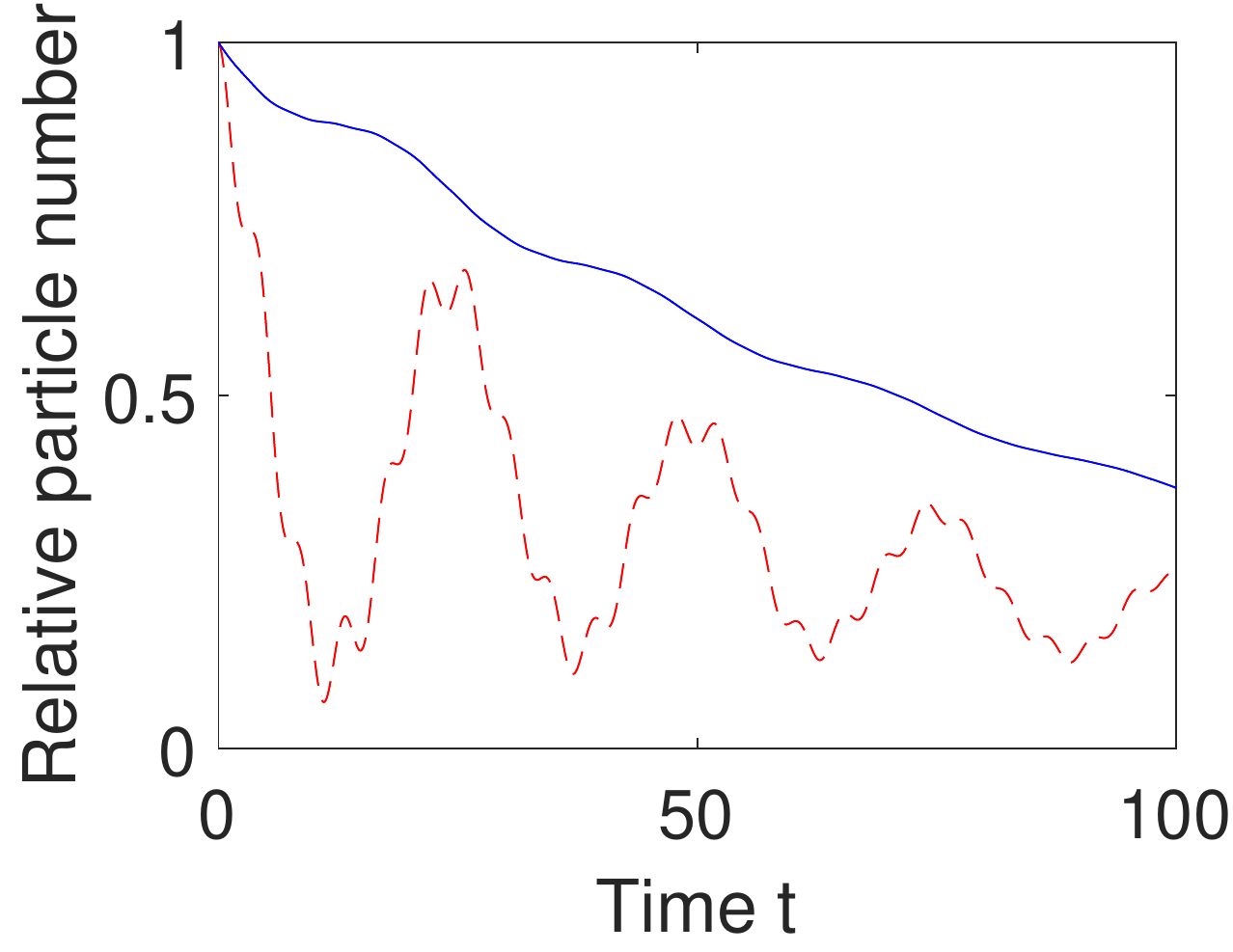}
\caption[ ]{\label{f-lindblad} Relative total particle number $(n_1(t)+n_2(t))/N$ (solid line) 
and relative particle number in the second well $n_2(t)/N$ (dashed line) as a function of time 
for an open Bose-Hubbard dimer with $\epsilon=0$, $v=0.3$, $c=0.6$, $N=2$ and decay 
with the rate $\gamma=0.02$ from the second well. Scaled units with $\hbar=1$ are used.}
\end{center}
\end{figure}
{\small 
\begin{listing}[1000]{1001}
N = 2; Np = N+1; Nout = 10; epsilon = 0; v = 0.3; c = 0.6;
a = diag(sqrt(1:N),1); ad = a'; I = eye(Np);          
a1 = kron(a,I); ad1 = a1';        
a2 = kron(I,a); ad2 = a2';         
H = epsilon*(ad1*a1-ad2*a2)+v*(ad1*a2+ad2*a1)+c/2*(ad1*a1-ad2*a2)^2; 
gamma = 0.02;
psi0_1 = zeros(Np,1); psi0_1(1) = 1;   
psi0_2 = zeros(Np,1); psi0_2(N+1) = 1; 
psi0 = kron(psi0_1,psi0_2);   
rho = psi0*psi0';   
dt = 0.05; tlist = (0:2000)*dt;   
n1 = zeros(1,length(tlist)); n2 = zeros(1,length(tlist));  
for l = 1:length(tlist)        
  n2(l) = trace(rho*ad2*a2); n1(l) = trace(rho*ad1*a1); 
  nk = 10;   
  for k = 1:10    
    rho_pred = rho-1i*(H*rho-rho*H)*dt/nk+0.5*gamma*(a2*rho*ad2...
	              -ad2*a2*rho+a2*rho*ad2-rho*ad2*a2)*dt/nk;
    rho_m = 0.5*(rho+rho_pred);
    rho = rho-1i*(H*rho_m-rho_m*H)*dt/nk+0.5*gamma*(a2*rho_m*ad2...
	      -ad2*a2*rho_m+a2*rho_m*ad2-rho_m*ad2*a2)*dt/nk;
  end;
end;
figure(1)  
hold on    
plot(tlist,n2/N,'r--'); plot(tlist,(n1+n2)/N,'b');
box on
xlabel('Time t'); ylabel('Relative particle number')
\end{listing}
}
\noindent
Figure \ref{f-lindblad} shows the resulting decay dynamics of the relative total particle 
number $(n_1(t)+n_2(t))/N$ (solid line) and the relative particle number in the second 
well $n_2(t)/N$ for a symmetric double well with $\epsilon=0$, tunneling coefficient 
$v=0.3$, interaction constant $c=0.6$, decay rate $\gamma=0.02$ and an initial particle 
number $N=2$. 

In a non-interacting open dimer, the occupation $n_2(t)/N$ would yield exponentially 
damped cosine-shaped Rabi oscillations. We clearly observe how this behavior is modified 
by the interaction between the particles. Various Fourier components occur which result 
from different excitation energies in the spectrum of the interacting bosonic system. 

\section{Concluding remarks}
\label{s-con}
Matrix representation techniques were presented and illustrated by a variety of different examples
demonstrating both their simplicity and wide applicability. These qualities make them suitable for 
use in research projects as well as quantum mechanics courses for undergraduate and graduate students.
\section*{Acknowledgments} The authors would like to thank Eva-Maria Graefe for
careful reading of the manuscript and for all
valuable comments and suggestions.

\section*{References}

\end{document}